# Transient Stability of Hybrid Power Systems Dominated by Different Types of Grid-Forming Devices

Xiuqiang He, *Student Member, IEEE*, Sisi Pan, and Hua Geng, *Fellow, IEEE*

*Abstract*—This paper investigates the transient stability of power systems co-dominated by different types of grid-forming (GFM) devices. Synchronous generators (SGs and VSGs) and droop-controlled inverters are typical GFM devices in modern power systems. SGs/VSGs are able to provide inertia while droop-controlled inverters are generally inertialess. The transient stability of power systems dominated by homogeneous GFM devices has been extensively studied. Regarding the hybrid system jointly dominated by heterogeneous GFM devices, the transient stability is rarely reported. This paper aims to fill this gap. It is found that the synchronization behavior of the hybrid system can be described by a second-order motion equation, resembling the swing equation of SGs. Moreover, two significant differences from conventional power systems are discovered. The first is that the droop control dramatically enhances the damping effect, greatly affecting the transient stability region. The second is that the frequency state variable exhibits a jump at the moment of fault disturbances, thus impacting the post-fault initial-state location and stability assessment. The underlying mechanism behind the two new characteristics is clarified and the impact on the transient stability performance is analyzed and verified. The findings provide new insights into transient stability of power systems hosting heterogeneous devices.

## I. Introduction

TRANSIENT stability matters for power system security. It refers to the ability of systems to remain in synchronism after being subjected to a large disturbance such as a short circuit or loss of generation. Transient stability characteristics of conventional power systems are mainly dominated by the motion of synchronous generators (SGs). In contrast, transient stability characteristics of power electronic inverters are primarily shaped by control strategies. Among various available control strategies, grid-forming (GFM) control and its variants have been considered a promising solution for low-inertia power systems [1]. In recent years, transient stability issues regarding GFM inverter-based systems have attracted wide attention [2], [3]. It was reported that different GFM controls lead to different transient stability characteristics [2]. Virtual synchronous generator (VSG) is a typical GFM scheme. By mimicking the motion behavior of SGs, VSG is designed to provide a grid-friendly inertial response. Droop control is another typical GFM scheme, the primary aim of which is to manage the synchronization and power-sharing among distributed generation units. More generally, SGs can also be considered naturally-existing GFM devices. A mixed deployment of various GFM resources is to be expected in future power systems as it facilitates the implementation of flexible and efficient stabilization and regulation functions [4], [5]. The Consortium for Electric Reliability Technology Solutions (CERTS) Microgrid is a real-world project case of such hybrid systems [5]. In the microgrid, cascading instability event under extreme disturbances has been reported [5]. This motivates us to theoretically investigate the transient stability of such hybrid systems. Unlike the system with the same type of devices, the hybrid system stability analysis is challenged by different GFM functionalities and their interactive behaviors.

Previously, transient stability of power systems hosting a single type of GFM device had been well documented in the literature. Specifically, the transient behavior of VSGs is inherited from SGs to a large degree [6]. The fundamental stability mechanism can be easily understood with the motion characteristics of SGs [6]–[9]. The transient instability is usually due to insufficient synchronizing torque, manifesting as first- or second-swing instability. Since the damping torque is generally negligible, the equal-area or energy-integral criterion was developed to give an intuitive insight [8]. Regarding droop-controlled systems, transient stability has also been reported [10]–[16]. The synchronization behavior of a single generation unit can be represented by a first-order equation [10]. Provided that there is a stable equilibrium point, viz., power flow on each branch is within its transfer limit, the synchronization can be guaranteed [16]. Also, a droop-controlled networked system was proved to be overdamped and therefore exponentially stable [16]. From existing studies, overall, it has been shown that in terms of transient stability mechanism there is a significant difference between VSG-controlled systems and droop-controlled ones. The difference results from the fact that droop control is generally inertialess, leading to an overdamped dynamic response [10]–[12].

Currently, regarding the hybrid system with heterogeneous GFM devices, very little is known about the transient stability mechanism. It remains unknown how to model, analyze, and understand the synchronization behavior of the system properly. It is also unclear how the hybrid system differs from or resembles conventional ones in certain aspects of stability characteristics. These issues cannot be addressed with current

This work was supported by the National Natural Science Foundation of China (U2066602, 52061635102). *(Corresponding author: Hua Geng.)*

X. He and H. Geng are with the Department of Automation, Beijing National Research Center for Information Science and Technology, Tsinghua University, Beijing, 100084, China (e-mail: hxq19@tsinghua.org.cn; genghua@tsinghua.edu.cn).

S. Pan is with the College of Electrical, Energy and Power Engineering, Yangzhou University, Yangzhou, 225000, China (e-mail: pss970503@icloud.com).

knowledge, and new insights into the hybrid system modeling and transient stability analysis should be provided.

This paper is aimed to fill this gap. The modeling of the hybrid system comprising an inertia-provided SG and an inertialess droop-controlled inverter is presented. It is found that the transient synchronization behavior of the hybrid system can be described by a second-order equivalent motion equation, analogous to the swing equation of an SG. Based on the equivalent motion equation, two significant differences between the hybrid system and the conventional one are found. Firstly, the damping effect of the hybrid system is dramatically enhanced by the droop control, producing a significant impact on the transient stability region. Note that in the conventional system, the mechanical friction damping effect of SGs is generally weak [17] and can be disregarded in transient stability assessment [8], [9]. Secondly, it is observed that the frequency governed in the droop-controlled subsystem exhibits a jump at the moment of fault disturbances. This is in fact due to its inertialess property. The frequency jump is followed by a jump of the post-fault initial state, thereby impacting the transient stability. Moreover, the essential distinction between the hybrid system and the droop-controlled one is identified. By comparatively assessing the transient stability boundary, dynamic performance, and fault critical clearing time (CCT), it is found that the hybrid system transient stability generally outperforms the SG-based one but underperforms the droop-controlled one.

The contributions of this paper are summarized as follows:
- A physically significant motion equation is derived for the hybrid system to analyze its transient synchronization behavior. The motion equation makes it possible to apply classical approaches from conventional power systems to transient stability studies of hybrid power systems.
- New important characteristics in the hybrid system distinguishing itself from conventional ones are elaborated. The impact of the enhanced damping and the frequency jump on the transient stability performance is clarified.
- Transient stability characteristics of the hybrid system, the SG-based conventional system, and the droop-controlled system are comparatively analyzed, and the similarities and differences between them are identified.

The findings provide a prospect to properly allocating and utilizing available control resources such as inertia and damping in future power systems. The rest of this paper proceeds as follows. The modeling of the hybrid system is presented in Section II. The stability mechanism of different types of systems is analyzed in Section III. The stability boundary and CCT are assessed and compared in Section IV. The simulation and experimental verifications are performed in Section V and the paper is concluded in Section VI.

## II. SYSTEM DESCRIPTION AND MODELING

A simple three-bus system is shown in Fig. 1, where two GFM devices and a constant-impedance load are included. The system is considered as a prototype of a power system comprising homogeneous or heterogeneous GFM devices, and also, it can represent transmission or distribution network systems. From the viewpoint of inertia provision, GFM devices can be roughly classified into two categories. The first category can provide inertia, e.g., SGs/VSGs. The simplified diagram of an SG is shown in Fig. 2(a). The motion equation is given by [18],

$$d\delta_1/dt = \omega_N(\omega_1 - \omega_0)$$
$$T_J d\omega_1/dt = P_1^* - P_1 - D_1(\omega_1 - \omega_0). \quad (1)$$

where $T_J$ denotes inertia time constant (also denoted by $H$, $T_J = 2H$ [19]), and $D_1$ denotes the damping coefficient. For VSGs, the damping is related to the droop characteristics. For SGs, the damping is related to the mechanical friction and the damping winding torque. The droop control (primary frequency regulation) of SGs is governed by a speed governor with a slow response [7].

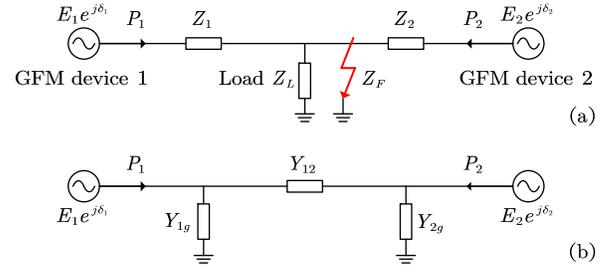

Fig. 1. (a) A prototype system comprising two GFM devices. (b) An equivalent circuit diagram to (a).

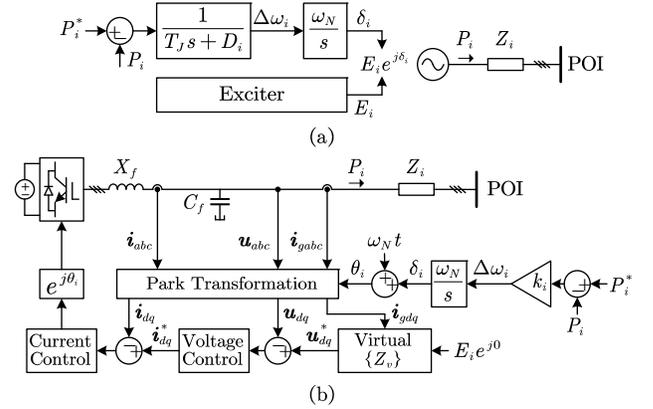

Fig. 2. (a) An SG considered as an inertia-provided GFM device. (b) A $P$-$f$ droop-controlled inverter considered as an inertia-less GFM device.

The second category of GFM devices is generally inertialess. The schematic diagram of a $P$-$f$ droop-controlled inverter is shown in Fig. 2(b). If a low-pass filter (LPF) is incorporated into the droop control loop, the inertia emulation can also be realized [19]. Typically, the bandwidth of the LPF is high, e.g., 20 Hz [20], which is much faster than the electromechanical synchronization behavior. Therefore, such unintentionally emulated inertia is ignorable, and the relationship between frequency and active power is still governed largely by the droop item [10]–[12]. The inertialess droop control is given by

$$d\delta_2/dt = \omega_N(\omega_2 - \omega_0) = \omega_N k_2(P_2^* - P_2). \quad (2)$$

where $k_2$ denotes the droop coefficient. The setting of its value is subject to the primary reserve for the maximum frequency deviation. To make the droop control work well, the inverter DC side should be able to provide power at the first instant after a disturbance and maintain the stability of the DC voltage. This can be achieved by power reserves and proper DC voltage

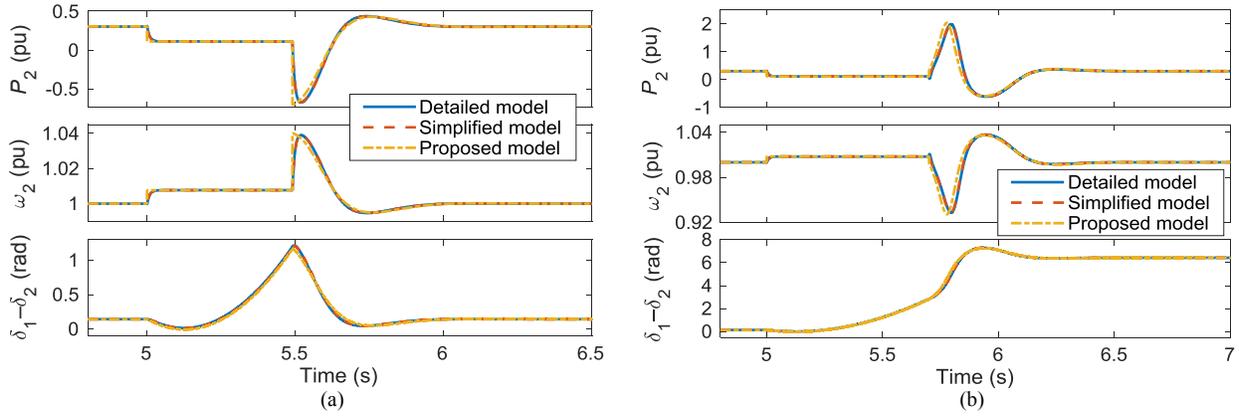

Fig. 3. Validations of the proposed model, where the detailed model incorporates the voltage and current dynamics as well as the LPF of the droop controller; the simplified model ignores the voltage and current dynamics but retains the LPF; the proposed model further ignores the LPF. In (a), transient stability remains after a grid fault, and the system recovers to the original equilibrium. In (b), the system slips to another equilibrium.

control [21].

To understand the synchronization behavior in the electromechanical time scale, simplified modeling can be applied. The dynamics of current and voltage control as well as the network dynamics are neglected [10]–[16], considering that these dynamics are much faster than the electromechanical time scale of interest. It should be noted in this study that transient stability is concerned with whether a system can remain in synchronism after a large disturbance, e.g., a short-circuit fault. The virtual impedance control can be applied to implement current limiting protection during the fault period [22]. Generally, the network voltage recovers rapidly after fault clearance, and consequently, the inverter device exits from the current limiting control [13]. For ease of analysis, the virtual impedance-based current limiting control is enabled only during the fault period in the modeling below.

Without loss of generality, suppose that GFM device 1 is the SG and GFM device 2 is the droop-controlled inverter. Denote by $Z_1$ the lumped impedance of the SG internal reactance with line impedance; denote by $Z_2$ the lumped impedance of the virtual impedance with line impedance. Eliminating the load bus, the equivalent circuit is obtained in Fig. 1(b), where the voltage angles are relative to the synchronous reference frame. The admittance in the equivalent circuit is given by,

$$1/Y_{12} = Z_1 + Z_2 + Z_1 Z_2/Z_3$$
$$1/Y_{1g} = Z_1 + Z_3 + Z_1 Z_3/Z_2 \quad (3)$$
$$1/Y_{2g} = Z_2 + Z_3 + Z_2 Z_3/Z_1$$

where $Z_3 = Z_L$ or $Z_L // Z_F$ in normal and short-circuit fault conditions, respectively.

The power output of the two sources can be written as

$$P_1 = E_1^2 (G_{12} + G_{1g}) - E_1 E_2 |Y_{12}| \cos(\delta_1 - \delta_2 + \gamma) \quad (4)$$
$$P_2 = E_2^2 (G_{12} + G_{2g}) - E_1 E_2 |Y_{12}| \cos(\delta_1 - \delta_2 - \gamma) \quad (5)$$

where $G_{12}$, $G_{1g}$, and $G_{2g}$ are the real parts of $Y_{12}$, $Y_{1g}$, and $Y_{2g}$, respectively, and $\gamma$ is the impedance angle of $Z_{12}$ corresponding to $Y_{12}$. The following two new state variables are defined to focus on the synchronization of the two sources,

$$\delta = \delta_1 - \delta_2, \quad \omega_e = \omega_1 - \omega_2. \quad (6)$$

Synthesizing (1) to (6) gives a second-order motion equation,

$$d\delta/dt = \omega_N \omega_e$$
$$T_{Jeq} d\omega_e/dt = P_M - P_E - D_{eq} \omega_e \quad (7)$$

where

$$P_M = P_1^* - E_1^2 (G_{12} + G_{1g}) - D_1 k_2 [P_2^* - E_2^2 (G_{12} + G_{2g})]$$
$$P_E = E_1 E_2 |Y_{12}| [D_1 k_2 \cos(\delta - \gamma) - \cos(\delta + \gamma)]$$
$$T_{Jeq} = T_J \quad (8)$$
$$D_{eq} = D_1 + k_2 T_J \omega_N E_1 E_2 |Y_{12}| \sin(\gamma - \delta).$$

From (7), it is important to find that the synchronization behavior of the hybrid system can be equivalently described by a physically significant swing equation. The equivalent inertia and damping coefficients are given in (8). It is interesting to find that the equivalent inertia remains unchanged but the equivalent damping is changed. Based on this finding, the transient stability mechanism can be readily understood with the concept of the swing equation.

Prior to using the model for stability analysis, it is verified to be able to capture the synchronization behavior of the system. The model is compared against a detailed model that contains the voltage and current controller, the network dynamics, and an LPF in the droop controller. A relatively simplified model is also considered, in which the LPF is retained with respect to the proposed model. The validation results are presented in Fig. 3. It is found that the accuracy of the model is hardly compromised when ignoring the voltage and current dynamics. With a 20 Hz bandwidth of LPF incorporated, the droop frequency shows a little inertial response. The inertia is ignorable in the electromechanical time scale. As seen from the result of the proposed model, the power angle response is quite close to the other two models. Regarding the validity of the proposed model, two remarks are drawn as follows. Firstly, the voltage and current controller parameters should be well tuned to make sure that these control loops are stable and respond quickly. Secondly, the droop controller is almost inertialess. If it is assigned non-ignorable inertia, the proposed model will probably fail. In this case, both the sources should be considered inertial [7].

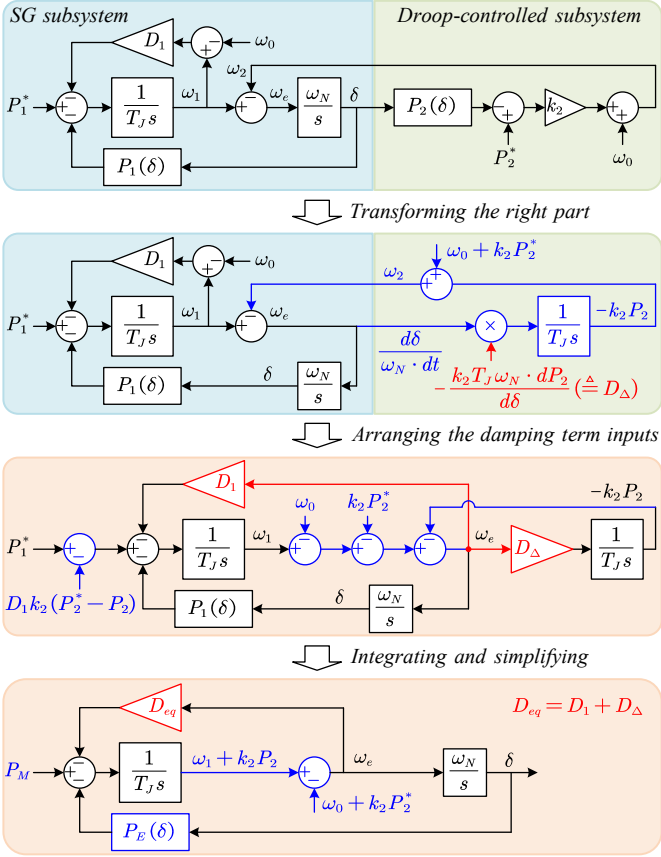

Fig. 4. Illustration of the droop control enhancing the system damping characteristics. Note that the state, $\omega_1 + k_2 P_2$, should be redefined at the moment of fault disturbances as $P_2$ jumps at this moment.

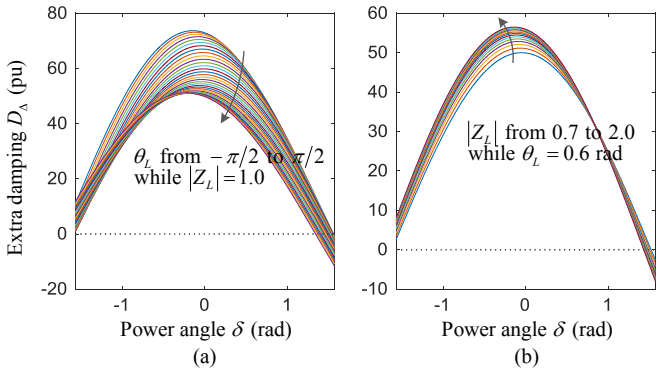

Fig. 5. The impact of the load impedance characteristics $Z_L = |Z_L|\angle\theta_L$ on the damping enhancement. (a) As $Z_L$ changes from capacitive to inductive, and damping enhancement effect weakens. (b) As the load becomes lighter, the damping enhancement effect increases.

## III. Transient Stability Mechanism Analysis

### A. Stability Mechanism Analysis of the Hybrid System

*1) Enhanced Damping Characteristics:* In a special case, if the droop gain $k_2$ is given zero, the output frequency of GFM device 2 will remain unchanged, and the device will be mathematically equivalent to an infinite bus. For this "single-machine infinite-bus (SMIB)" system, the motion equation of GFM device 1 is a special case of (7), as follows,

$$d\delta_1/dt = \omega_N(\omega_1 - \omega_0)$$
$$T_J d\omega_1/dt = P_M - P_E - D_1(\omega_1 - \omega_0) \quad (9)$$

where $P_M = P_1^* - E_1^2(G_{12} + G_{1g})$, $P_E = -E_1 E_2 |Y_{12}|\cos(\delta + \gamma)$.

Comparing (8) with (9), it is important to note that the equivalent inertia in the hybrid system remains the same as the generator inertia whereas the equivalent damping is enhanced by a variable term. Physically, the inertialess GFM control cannot provide inertia emulation, and therefore the overall inertia remains unchanged. Owing to the droop control, however, the inertialess GFM inverter itself is overdamped [10], and consequently, the overall damping of the hybrid system is significantly enhanced. The damping term can produce energy dissipation in dynamic processes, therefore affecting the stability and dynamic performance. A considerable energy dissipation, on one hand, has an important impact on the transient stability boundary [23]. On the other hand, it can improve dynamic performance. This is the first significant difference of the hybrid system from the conventional one. In the latter, the mechanical friction damping effect is relatively weak [17] and can be ignored in transient stability assessment (TSA) [8], [9]. Note that damping has also an important impact on small-signal stability, which is outside the focus of this study.

To facilitate an intuitive understanding of the enhanced damping characteristics by the placement of droop control, the model transformation is displayed in Fig. 4. At first, a differential and then integral operation is performed on the droop control term. By doing this, the gain coefficient $D_\Delta$ of the damping additionally introduced by the droop control is formulated. Then, the model is arranged so that the SG damping term and the additional damping term have the same input. Last, the two damping terms are merged and the model description is simplified, giving the equation describing the synchronization behavior. From Fig. 4, it can be confirmed again that an additional damping effect can be introduced by the droop control.

In (8), it is identified that the magnitude of the damping gain is directly related to the droop gain $k_2$ and the connection strength $|Y_{12}|$. The larger the droop gain or the closer the connection between the two devices, the stronger the damping effect introduced. Based on the expression of the additional damping term, a few remarks can be drawn regarding the impact of the transmission line and load impedance characteristics on the damping enhancement. Firstly, the larger the transmission line impedance $|Z_1|, |Z_2|$, the looser the connection and therefore the weaker the damping enhancement. Secondly, with the load impedance changes from capacitive to inductive, $|Z_{12}|$ basically increases and accordingly $|Y_{12}|$ decreases, as a result of which, the damping enhancement effect becomes weak. This is shown in Fig. 5(a). Thirdly, the impact of the load size ($|Z_L|$) varies with the load impedance angle. For an inductive load, the damping enhancement will increase if the load becomes light, as exemplified in Fig. 5(b).

*2) Frequency Jump Characteristics:* It is seen in Fig. 4 that the frequency state variable $\omega_e$ can be expressed by,

$$\omega_e = \omega_1 + k_2 P_2 - \omega_0 - k_2 P_2^* \quad (10)$$

which can also be directly obtained through (2) and (6). When a network fault occurs or is cleared, the active power $P_2$ changes abruptly, leading to a jump of the frequency $\omega_e$. This is the case in the droop-controlled subsystem as the filtering time constant is generally small and negligible [10]–[12]. For the conventional power system, this is not the case because the change of frequency is restricted by the moment of inertia. Thus, it is identified that the jump in the frequency state variable is the second significant point that distinguishes the hybrid system from the conventional power system.

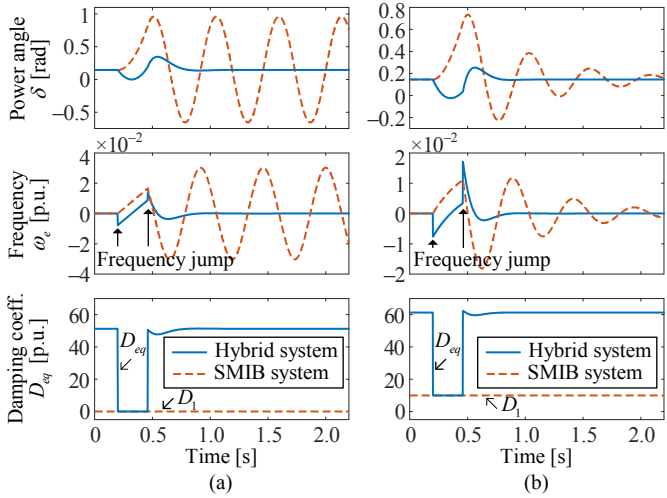

Fig. 6. Comparisons of the transient response between the hybrid system and the SMIB system. (a) The SG damping coefficient $D_1 = 0$. (b) $D_1 = 10$. The other system parameters are given in Table I.

TABLE I
SYSTEM PARAMETERS FOR A HIGH-VOLTAGE NETWORK CASE

| Symbol | Description | Value |
| --- | --- | --- |
| $U_N$ | Voltage level | 110 kV |
| $S_N$ | Nominal capacity | 100 MVA |
| $\omega_N$ | Nominal frequency | $100\pi$ rad/s |
| $E_1, E_2$ | Internal voltage | 1.1, 1.1 p.u. |
| $P_1^*, P_2^*$ | Mechanical power or power reference | 0.5, 0.3 p.u. |
| $Z_L$ | Load impedance | 0.82 + j0.57 p.u. |
| $Z_1$ | Lumped impedance | 0.05 + j0.44 p.u. |
| $Z_2$ | Lumped impedance | 0.10 + j0.30 p.u. |
| $Z_v$ | Virtual impedance | 0 + j0.75 p.u. |
| $Z_F$ | Fault impedance | 0.01 Ω |

Whether a nonlinear system is stable depends on not only the stability boundary but also the initial state. The frequency jump characteristics should be paid attention to in TSA as it affects the determination of the post-fault initial state. Denote by the subscript "0−" the moment before fault clearance and "0+" the moment after fault clearance. Considering that the state trajectory during the fault can be obtained from simulations [6], the power angle $\delta_{0^-}$ and the SG frequency $\omega_{1,0^-}$ are known. At the moment after fault clearance, the initial frequency becomes

$$\omega_{e,0^+} = \omega_{1,0^-} + k_2 P_{2,0^+} - \omega_0 - k_2 P_2^* \quad (11)$$

where $P_{2,0^+}$ represents the inverter output power,

$$P_{2,0^+} = E_2^2(G_{12} + G_{2g}) - E_1 E_2 |Y_{12}| \cos(\delta_{0^-} - \gamma). \quad (12)$$

As discussed above, the enhanced damping and the frequency jump are identified as new characteristics in the synchronization behavior of the hybrid system. Considering two cases below to verify the characteristics:

i. Case A: $k_2 = 0.04$, $D_1 = 0$ versus $k_2 = 0$, $D_1 = 0$;
ii. Case B: $k_2 = 0.04$, $D_1 = 10$ versus $k_2 = 0$, $D_1 = 10$;

where, for ease of comparison, $k_2$ is given zero to yield an "SMIB" system as a comparison object. It should be also noted that $D_1 = 0$ is an ideal case only for SGs. For VSGs, it is improper to zero $D_1$ as it should reflect the droop characteristics.

In Fig. 6, it is seen that the hybrid system converges significantly faster than the SMIB system. In Fig. 6(a), the SMIB system is undamped, exhibiting a constant-amplitude oscillation after the disturbance. Benefiting from the additional damping effect, the hybrid system settles down rapidly. The damping coefficient curve in Fig. 6 indicates that the pre-fault damping and the post-fault damping are enhanced whereas the damping during the fault period remains almost unchanged. This is because the grounding impedance reduces $|Y_{12}|$ sharply, weakening the connection between the droop control and the SG. In Fig. 6(b), it is shown that even if the SG's damping is large, the overall damping effect can still be improved. Additionally, it is observed in Fig. 6 that the hybrid system frequency changes abruptly at the moment of disturbances, As will be shown in Section IV, the frequency value after the change should be of concern in TSA.

### B. Discussions on More GFM Devices and Load Types

In a hybrid system with multiple GFM devices, the damping enhancement effect and the frequency jump still happen. The latter is apparent because of the inertialess property of droop control. Considering a multi-inverter multi-generator system, droop control enhancing the entire damping effect is demonstrated as follows. Suppose there are $m$ synchronous generators and $n$ droop-controlled inverters in the system. The motion of the generators is governed by (ignoring the damping term for simplicity)

$$\begin{cases} \dot{\delta}_i = \omega_N(\omega_i - \omega_0) \\ T_{J_i}\dot{\omega}_i = P_i^* - P_i \end{cases}, i = 1, 2, \cdots, m. \quad (13)$$

The motion of the inverters is governed by

$$\omega_j = \omega_0 + k_j(P_j^* - P_j), \; j = m+1, m+2, \cdots, m+n. \quad (14)$$

The well-known concept of center of angle (COA) [24] is leveraged to define the collective motion of the generators as a whole during a transient response. The angle and angular velocity of the COA can be formulated by [24],

$$\delta_{\text{COA1}} = \frac{1}{T_{\text{COA1}}}\sum_{i=1}^{m} T_{J_i}\delta_i, \; \omega_{\text{COA1}} = \frac{1}{T_{\text{COA1}}}\sum_{i=1}^{m} T_{J_i}\omega_i, \\ T_{\text{COA1}} = \sum_{i=1}^{m} T_{J_i} \quad (15)$$

which gives rise to the equation governing the COA of the generators as follows [24],

$$\begin{cases} \dot{\delta}_{\text{COA1}} = \omega_N(\omega_{\text{COA1}} - \omega_0) \\ T_{\text{COA1}}\dot{\omega}_{\text{COA1}} = \sum_{i=1}^{m}(P_i^* - P_i). \end{cases} \quad (16)$$

Further, the collective motion of the inverters is defined by,

$$\delta_{\text{COA2}} = k_{\text{COA2}} \sum_{j=m+1}^{m+n} \frac{\delta_j}{k_j}, \quad \omega_{\text{COA2}} = k_{\text{COA2}} \sum_{j=m+1}^{m+n} \frac{\omega_j}{k_j},$$
$$\frac{1}{k_{\text{COA2}}} = \sum_{j=m+1}^{m+n} \frac{1}{k_j} \quad (17)$$

which gives the equation governing the COA of the inverters,

$$\begin{cases} \dot{\delta}_{\text{COA2}} = \omega_N (\omega_{\text{COA2}} - \omega_0) \\ \omega_{\text{COA2}} = \omega_0 + k_{\text{COA2}} \sum_{j=m+1}^{m+n} (P_j^* - P_j). \end{cases} \quad (18)$$

Considering the damping effect of the COA of the inverters on that of the generators, the relative motion between them is defined by $\tilde{\omega}_{\text{COA}} = \omega_{\text{COA1}} - \omega_{\text{COA2}}$.

$$T_{\text{COA1}} \dot{\tilde{\omega}}_{\text{COA}} = \sum_{i=1}^{m} (P_i^* - P_i) + T_{\text{COA1}} k_{\text{COA2}} \sum_{j=m+1}^{m+n} \dot{P}_j \quad (19)$$

where $P_j = \sum_{i=1}^{m+n} E_j E_i |Y_{ji}| \cos(\delta_j - \delta_i + \gamma_{ji})$ represents the real power of the $j$th inverter ($|Y_{ji}| \angle \gamma_{ji}$ is the element of node admittance matrix). It can be obtained further that

$$T_{\text{COA1}} \dot{\tilde{\omega}}_{\text{COA}} = \sum_{i=1}^{m} (P_i^* - P_i)$$
$$- T_{\text{COA1}} k_{\text{COA2}} \sum_{j=m+1}^{m+n} \sum_{i=1}^{m+n} E_j E_i |Y_{ji}| \omega_N \sin(\delta_j - \delta_i + \gamma_{ji})(\omega_j - \omega_i) \quad (20)$$

Considering the following simplifications to yield an intuitive understanding of the damping enhancement. Assuming that the network line is inductive ($\gamma_{ji} = -\pi/2, j \neq i$), Equation (20) can be simplified as follows,

$$T_{\text{COA1}} \dot{\tilde{\omega}}_{\text{COA}} = \sum_{i=1}^{m} (P_i^* - P_i) - \sum_{j=m+1}^{m+n} \sum_{i=1}^{m} D_{ji}(\omega_i - \omega_j) \quad (21)$$

where $D_{ji} = T_{\text{COA1}} k_{\text{COA2}} E_j E_i |Y_{ji}| \omega_N \cos(\delta_j - \delta_i)$. It is indicated by (21) that the angular velocity difference between any generator and inverter can contribute damping to the relative motion between the COA of the generators and that of the inverters. This therefore briefly demonstrates the damping enhancement effect by droop control. It can also be found that the extra damping in the simple system [see (8)] is a special case of (21). To analyze how each droop-controlled inverter contributes to enhancing the damping of the generators, the network topology and its impedance characteristics should be considered. The network can be represented by a node admittance matrix, by which the electromagnetic power of each inverter can be derived. Physically, through the power exchange between the inverter and the network and further the generators, the damping in the transient response of the generators can therefore be enhanced.

Regarding the load type, the simplest one is constant impedance loads. If simply ignoring the amplitude variation of the load bus voltage in the post-fault response, a PQ load will be equivalent to a constant-impedance load. Moreover, considering constant-current loads, it can also be verified that droop control can enhance the damping of synchronization behaviors. Briefly, considering the current absorption at the load bus, deriving the output power of droop-controlled inverters, and then substituting it into (19) as above can give the additional damping term.

## C. Stability Characteristics Differences Between Different Types of Systems

It is of importance to further identify the differences in stability characteristics between the hybrid system and the system comprising homogeneous GFM devices. A system comprising two SGs is considered at first. Supposing uniform damping condition, $D_1/T_{J1} = D_2/T_{J2}$, the swing equation representing the two-generator synchronization behavior is given by [9],

$$\begin{aligned} d\delta/dt &= \omega_N \omega_e \\ T_{Jeq} d\omega_e/dt &= P_M - P_E - D_{eq} \omega_e \end{aligned} \quad (22)$$

where

$$T_{Jeq} = T_{J1} T_{J2}/(T_{J1} + T_{J2})$$

$$P_M = \frac{T_{J2} \left[ P_1^* - E_1^2 (G_{12} + G_{1g}) \right] - T_{J1} \left[ P_2^* - E_2^2 (G_{12} + G_{2g}) \right]}{T_{J1} + T_{J2}}$$

$$P_E = \frac{-T_{J2} E_1 E_2 |Y_{12}| \cos(\delta + \gamma) + T_{J1} E_1 E_2 |Y_{12}| \cos(\delta - \gamma)}{T_{J1} + T_{J2}}$$

$$D_{eq} = D_1 T_{J2}/(T_{J1} + T_{J2}).$$

As indicated by (22), the two-generator system and the SMIB system follow a similar transient stability mechanism. This is actually a basic concept of the extended equal-area criterion (EEAC) [9].

For a system comprising two inertialess droop-controlled inverters, the P-f droop control law is given as,

$$d\delta_i/dt = \omega_N k_i (P_i^* - P_i), \, i = 1, 2. \quad (23)$$

Denote the included angle between the two sources by,

$$\delta = \delta_1 - \delta_2. \quad (24)$$

From (4), (5), (23), and (24), the first-order motion equation representing the synchronization behavior is derived as [11],

$$d\delta/dt = A + B\cos\delta + C\sin\delta \quad (25)$$

where

$$A = \omega_N k_1 (P_1^* - E_1^2 G_{12} - E_1^2 G_{1g}) - \omega_N k_2 (P_2^* - E_2^2 G_{12} - E_2^2 G_{2g})$$
$$B = \omega_N E_1 E_2 G_{12} (k_1 - k_2)$$
$$C = \omega_N E_1 E_2 B_{12} (k_1 + k_2).$$

It has been indicated in [11] that $A^2 \leq B^2 + C^2$ is a sufficient and necessary condition for the existence of a stable equilibrium point as well as for asymptotic stability. The physical meaning behind this condition is that the power reference and the droop gain given should meet the power transfer limit requirement of the transmission line $Y_{12}$ [see Fig. 1(b)]. With this condition, the system can also be theoretically proved exponentially stable [16].

Regarding the differences in stability characteristics of different types of systems, a few concluding remarks can be drawn as follows. Mathematically, the description of the hybrid system synchronization behavior is similar to the conventional power system. Both can be represented with a second-order motion equation. Physically, significant energy dissipation due to the enhanced damping effect is observed in the hybrid system during the post-fault first swing. This notably distinguishes the hybrid system from the conventional one. In the latter, the energy dissipation during the first swing is minimal. If the first swing is stable, then the hybrid system will converge rapidly.

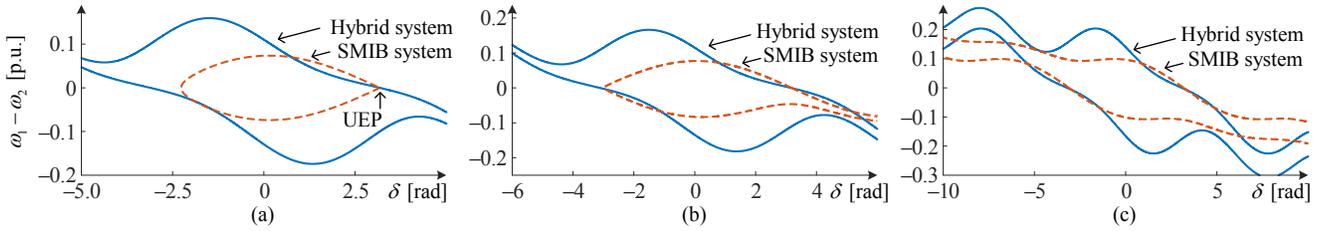

Fig. 7. The transient stability region is changed by the significant damping effect. (a) $D_1 = 0$. (b) $D_1 = 1.5$. (c) $D_1 = 10$. Note that the stability region of the SMIB system is not covered fully by that of the hybrid system, although the latter one has a more larger area than the former.

Oscillation instability due to insufficient damping is unlikely to occur. Another difference is the frequency jump. The state after the jump should be taken as the initial state in TSA. Moreover, there is an essential difference between the hybrid system and the droop-controlled one. This is because the hybrid system is still equipped with part of inertia, which leads to a typical second-order dynamic response, viz., swing and overshooting. For the droop-controlled system, if there is a steady-state operating point, the system state will directly tend to the steady state.

## IV. Transient Stability Assessment and Comparisons

For different types of systems, TSA is performed separately in this section. Additionally, the differences in transient stability boundary and CCT are comparatively investigated.

### A. Transient Stability Assessment (TSA) Approach

Regarding the second-order motion equation, almost all direct methods for TSA (mostly energy function-based methods) suffer from a certain degree of conservatism due to the energy dissipation of the damping term [23]. In contrast, by resorting to numerical methods, it is possible to obtain a less conservative stability region. In this regard, an inverse time integral-based approach can be applied to accurately identify the transient stability boundary [25]. Beginning from the unstable equilibrium point (UEP) and along the direction of the stable eigenvector to perform integral backward, the state trajectory obtained can represent the stability boundary. Regarding the first-order equation (25), an analytical approach can be used to directly give a one-dimensional stability boundary [10]–[12]. Regarding the consideration of the load type in TSA by direct methods, a great majority of previous research efforts dealt with only constant-impedance loads. There have also been a few advanced methods to consider nonlinear loads, such as the Ward equivalencing method employed in [26], [27] and the structure-preserving method initiated in [28]. These methods developed for conventional power systems can potentially lend themselves to the hybrid power system for stability assessment. This will be a challenge in future research.

The hybrid system stability boundary under different parameters is given in Fig. 7. A few general remarks can be drawn at first by observing Fig. 7. With a zero damping coefficient, the stable boundary is represented by a closed curve, which also represents the energy surface crossing through the UEP. If the damping is improved slightly, the stability region will be expanded, where the lower right corner opens, as displayed by the dashed line in Fig. 7(b). With large damping, the stability region is expanded further, and both the upper left corner and the lower right one become open, as seen in Fig. 7(c). For the hybrid system, owing to the damping enhanced, both the corners of the stability region are always open. Observing Fig. 7, it is also found that if the SG damping is small originally, the stability boundary can be greatly improved by the droop control. If it is not small, however, the improvement will be insignificant.

Only a single-period stable region is plotted in Fig. 7. Stability regions in the $2\pi$ periods are mathematically equivalent to one another, but there is an important difference in engineering. This is because pole slipping occurs while moving from one region to another [29]. Generally, only the stability region in the present period is of concern.

### B. Transient Stability Boundary of the Hybrid System

Considering $k_2 = 0.04$ for the droop control and $T_{J1} = 3$, $D_1 = 1.5$ for the SG as an example, the power-angle curve, the transient stability boundary, and the transient response of the hybrid system are plotted in Fig. 8(a). It is seen that the stability region is open rather than closed due to the strong damping effect. Additionally, since the frequency jump at the time of fault clearance, the critical clearing angle (CCA) is no longer the point from which the fault-on trajectory crosses the stability boundary. Instead, the critically stable scenario corresponds to that the post-fault initial state falls on the stable boundary. This is illustrated in Fig. 8(a) by the intersection of the blue dashed line with the stable boundary.

### C. Comparison With the SG-Based System

Considering a SMIB system at first, which is equivalently represented by $k_2 = 0$. The SG parameters $T_{J1} = 3$, $D_1 = 1.5$ remain the same as the hybrid system. The power angle curve and the stability boundary are plotted in Fig. 8(b). In terms of the power angle curve, there is a little different from that in Fig. 8(a) as $D_1 k_2$ is small and negligible for the hybrid system. Since the SMIB system damping is small, the upper left corner of the stability region is closed, which leads to a small stability region. Moreover, the CCAs for the two systems are similar. Since the frequency jump occurs in the hybrid system, however, the post-fault trajectory is affected and the resultant CCT is relatively large. It should be noted that the difference in the CCT is not a general conclusion as it is also affected by other parameters.

Then a two-generator system is considered where $T_{J1} = 3$, $D_1 = 1.5$, $T_{J2} = 3$, and $D_2 = 1.5$. As seen in Fig. 8(c), $P_M$ becomes so large during the fault period, leading to an over-much acceleration and consequently a smaller CCA than that for the SMIB system. Additionally, since the equivalent inertia is reduced by half [see (22)], the CCT becomes smaller.

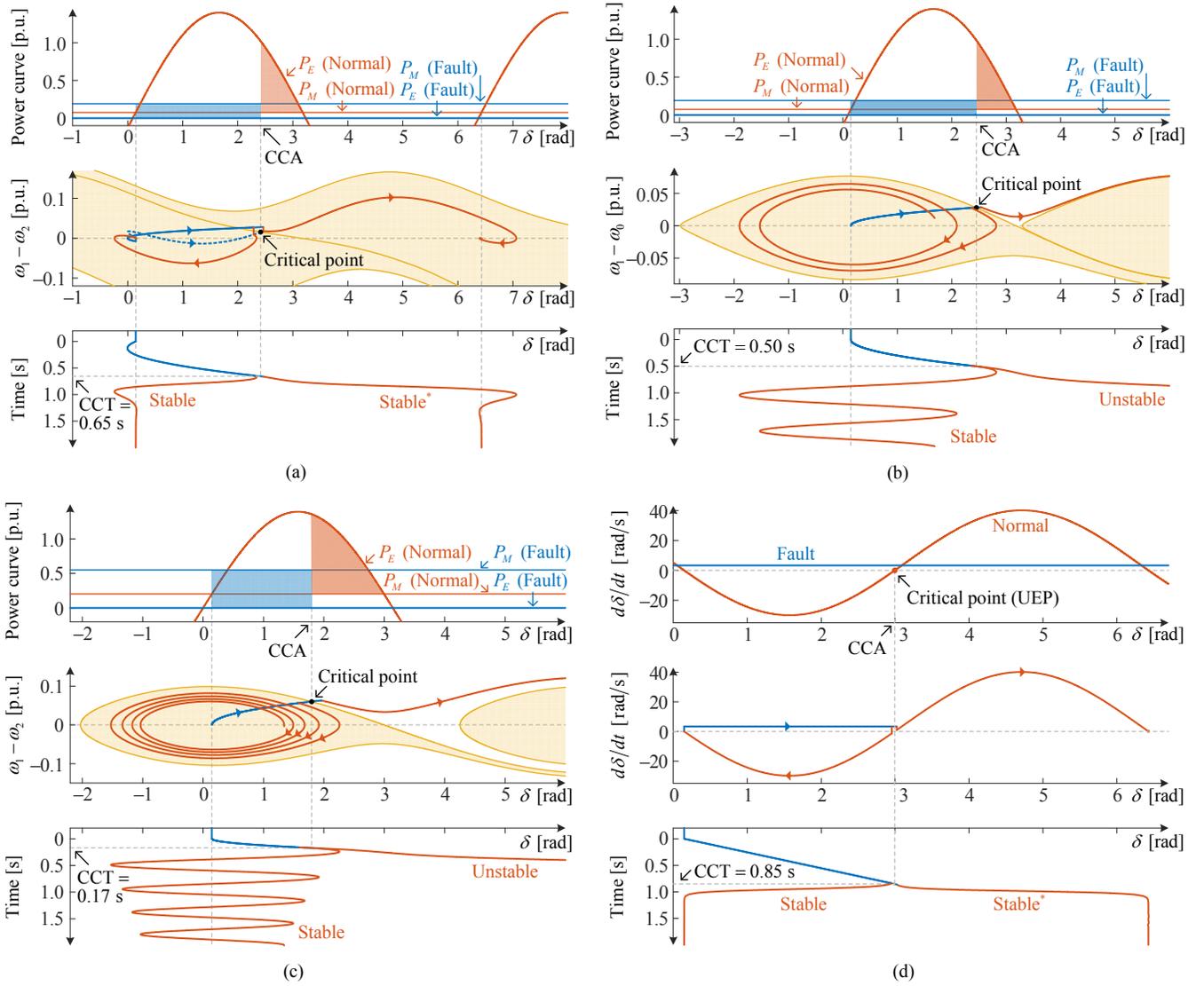

Fig. 8. Power angle curve, transient stability boundary, and transient response. (a) Hybrid system. (b) SIMB system. (c) Two-generator system. (d) Two-inverter droop-controlled system. On the phase plane in the subfigure (a), the blue dotted line represents the initial-state location after the fault is cleared at any time. The line is different from the blue solid line (i.e., the fault-on trajectory) due to the frequency jump. "Stable*" refers to mathematically stable but it is considered unstable in engineering as pole slipping occurs [29].

Through these comparisons, it is indicated that the hybrid system stability region is wider and the dynamic performance is also better than the SG-based system. These advantages benefit from the fact that the damping characteristics are greatly enhanced while the inertia characteristics remain. However, the improvement of CCT is inconclusive. For one thing, the fault-on trajectory always tends to the upper right corner on the phase plane. Unfortunately, this corner of the stability region is not expanded (see Fig. 7). For another, the frequency jump may worsen the CCA in some cases.

### D. Comparison With the Droop-Controlled System

Based on the first-order equation (25), the phase portrait can be straightforwardly applied to describe the synchronization behavior. As shown in Fig. 8(d), there are no equilibrium points during the fault period, and therefore the power angle keeps increasing. The CCA is determined by the UEP [10]. If the fault clearing time is beyond the corresponding CCT, the system will converge to the steady-state point in the next period. Since the system is overdamped, there is no overshooting.

## V. SIMULATION AND EXPERIMENTAL VERIFICATIONS

### A. Verifications on the Prototype Systems

The focus in the verifications is on the CCT and dynamic performance for different types of systems. A long-lasting and bolted fault is simulated to fully verify the CCT. The system parameters are provided in Table I. The simulation results are gathered in Fig. 9. It can be observed that the hybrid system outperforms the SG-based system in both transient stability and dynamic performance, but it is inferior to the droop-controlled system. In Fig. 9(a), the frequency jump can be observed. The droop-controlled system exhibits similar phenomena in Fig. 9(d)

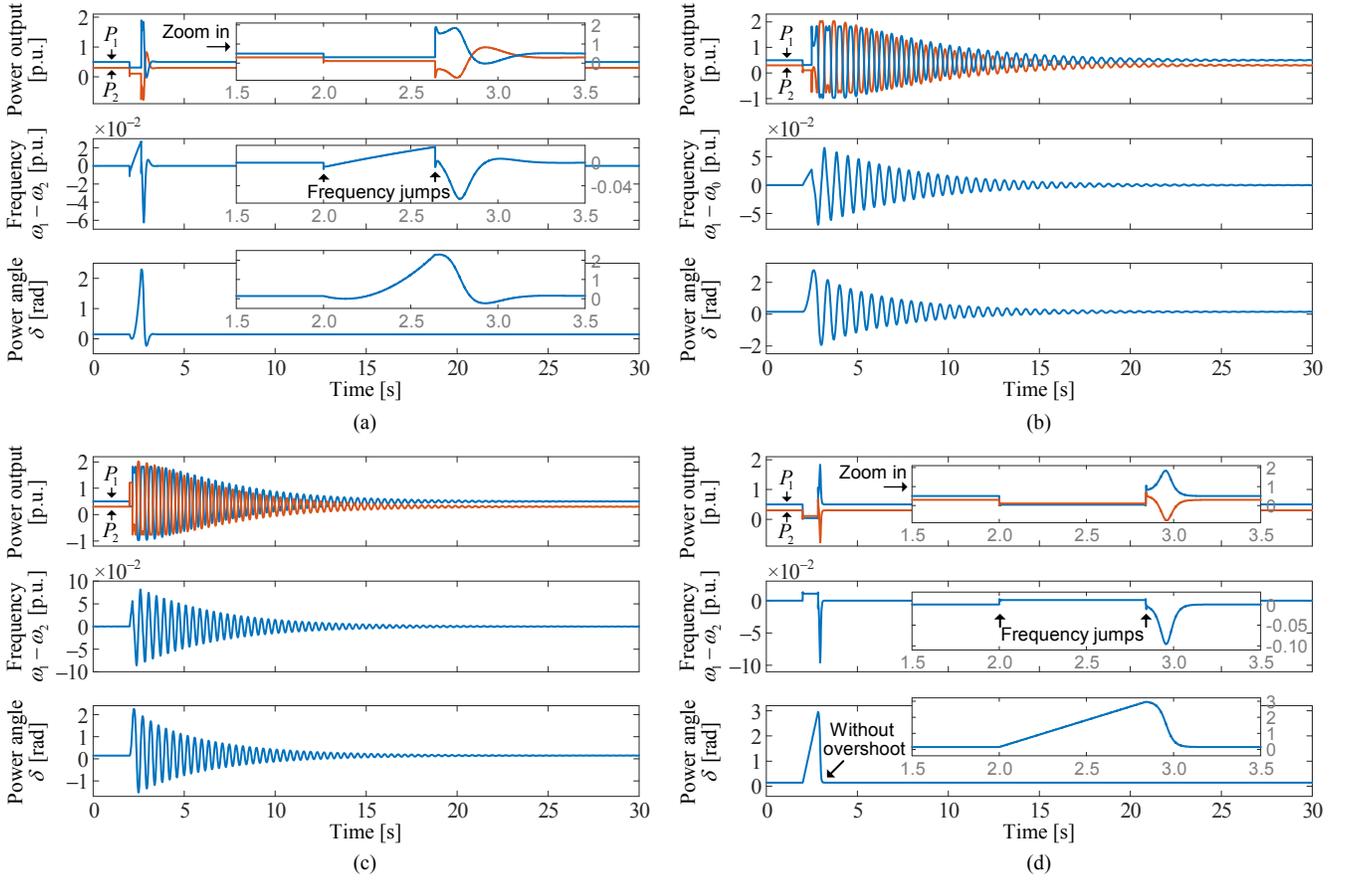

Fig. 9. Simulation verifications of CCT and dynamic performance for different types of systems. (a) In the hybrid system, the fault lasts 0.64 s. (b) In the SMIB system, the fault lasts 0.49 s. (c) In the two-generator system, the fault lasts 0.16 s. (d) In the two-inverter droop-controlled system, the fault lasts 0.84 s.

due to its inertialess property.

The transient synchronization behavior of the hybrid system is further verified by hardware-in-the-loop (HIL) experiments. The HIL experimental setup is shown in Fig. 10. The inverter is represented by an average-value model. The voltage-source interface, the LC filter, the network, and the generator are simulated in a SpaceR-Vx simulator. It runs in real time with a fixed 2e–5 seconds step size. The inverter is controlled by a dSPACE control board. The sampling and control period of it is fixed 1e–4 seconds. The experimental data is recorded by the host application of dSPACE. The parameters for experiments can be found in Table I. The experimental results are presented in Fig. 11, in which two cases are shown. A stable case is shown in Fig. 11(a), which coincides with the analysis in Fig. 8(a) and the simulation result in 9(a). In the case of Fig. 11(b), the system stabilizes at the adjacent equilibrium point, which agrees with the analysis in Fig. 8(a). Moreover, both the strongly damped dynamic performance and the droop frequency jumps are observed in the experimental results.

Note that the inverter output power reaches close to 2 p.u. in the post-fault transient response. In this case, the current limiting control should be enabled again to protect the device. If the virtual impedance scheme is adopted, the model and analysis in this study will apply. The transient stability region is expected to shrink as the virtual impedance leads to a weaker connection [6]. If the saturation limiting is adopted instead, a novel nonlinear stability analysis will be required [30], [31]. This issue should be further studied in the future, particularly for the systems comprising heterogeneous GFM devices.

### B. Verifications on the WSCC 9-Bus System

The findings are further verified on the WSCC 9-bus system. In Fig. 12, the SG at bus 3 is replaced with a droop-controlled inverter. Simulating a bolted fault at bus 9, the CCTs for the original system and the hybrid one are identified as 0.31 s and 0.90 s, respectively, by increasing the fault clearing time in a 0.01 s step. The simulation and experimental results are presented in Figs. 13 and 14, respectively. It is seen that the results are similar and both confirm the findings regarding damping enhanced and frequency jump change.

It is important to note that the hybrid system outperforms the original generator-based system in terms of CCT in this verification case. However, this is not a general remark. If the fault occurs at other buses, the hybrid system probably underperforms the original system in the aspect of CCT. For example, it has been verified that in the case of the fault occurring at bus 8, the CCT decreases from 0.51 s to 0.44 s. Referring to the analysis before, the CCT decreasing is because the hybrid system stability region cannot fully cover the original one (see Fig. 7) and due to the frequency jump change affecting the CCA and CCT [see Fig. 8(a)].

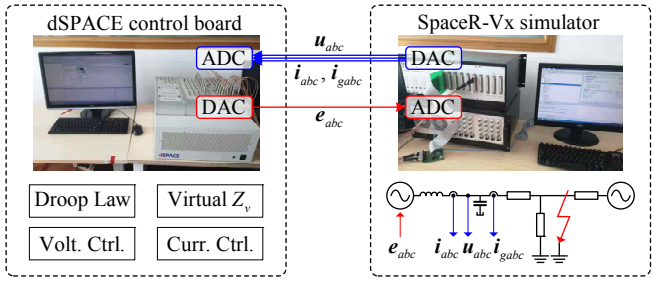

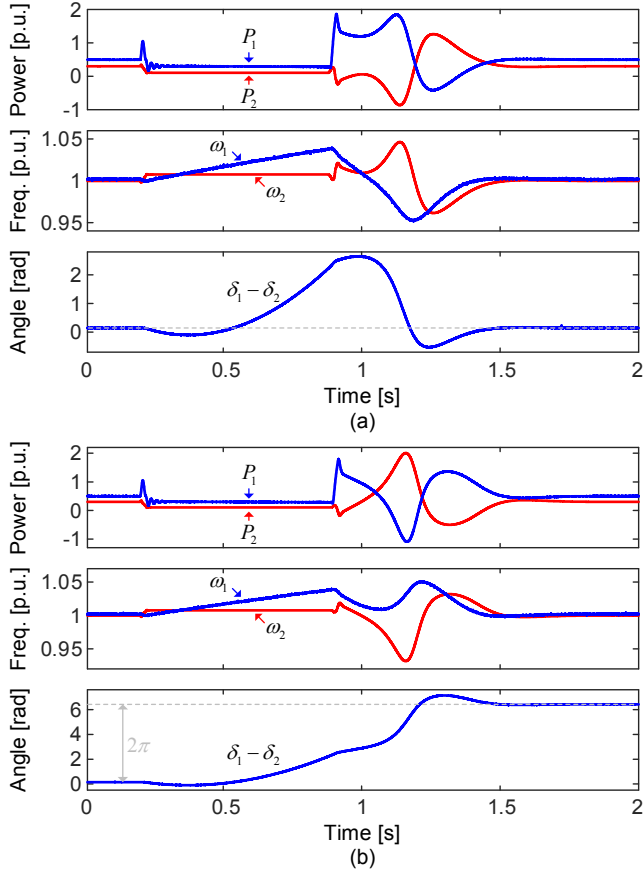

Fig. 10. The HIL experimental setup, where the dSPACE control board plays the role of controller whereas the SpaceR-Vx simulator simulates the inverter interface and the rest of the system.

Fig. 11. Experimental results of the two-source hybrid system. (a) The system remains in synchronism after a fault lasting 0.68s. The result resembles the simulation result in Fig. 9(a) (the theoretically predicted CCT is 0.65s). (b) The system converges to another equilibrium after a fault lasting 0.69s.

### C. Discussions

Though droop control can provide damping, the droop frequency jump is not friendly to frequency stability performance in terms of traditional power systems. Synchronous generators cannot withstand a high rate of change of frequency (RoCoF) because of mechanical limitations [32]. However, in an all-inverter microgrid where there are no synchronous generators, a high RoCoF is probably acceptable as long as the system is predicted to be stable under disturbances in advance. On the flip side, it is important to note that a rapid frequency change poses challenges for fault detection and relay protection. Additionally, the frequency regulation function such as fast fre-quency response has to act quickly to adapt to the rapid response of the system frequency. Shortly, the role of inertia in future power systems needs to be comprehensively considered. Moreover, considering the impact of inertia emulation and droop control on transient stability, frequency stability, oscillations, relay protection, etc., how to flexibly configure and manage available GFM resources needs more research [2], [33], [34].

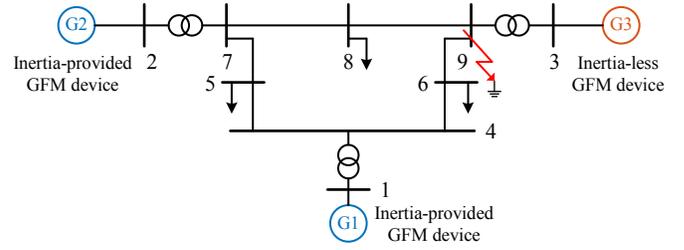

Fig. 12. G3 in the WSCC system is replaced with a droop-controlled inverter.

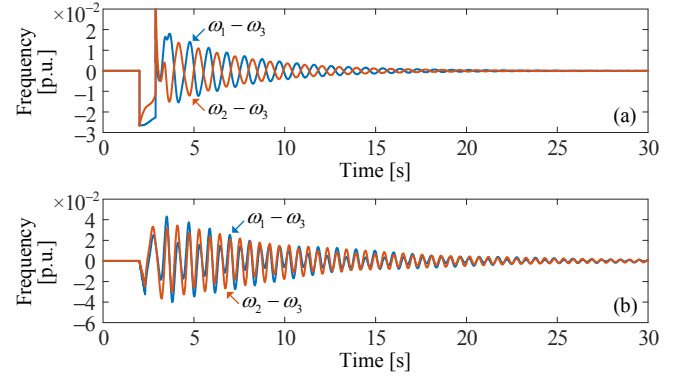

Fig. 13. Simulation result of the bus-9 fault. (a) In the hybrid system, the fault lasts 0.90 s. (b) In the WSCC system, the fault can last 0.31 s at most. Enhanced damping and frequency jump are observed in the hybrid system.

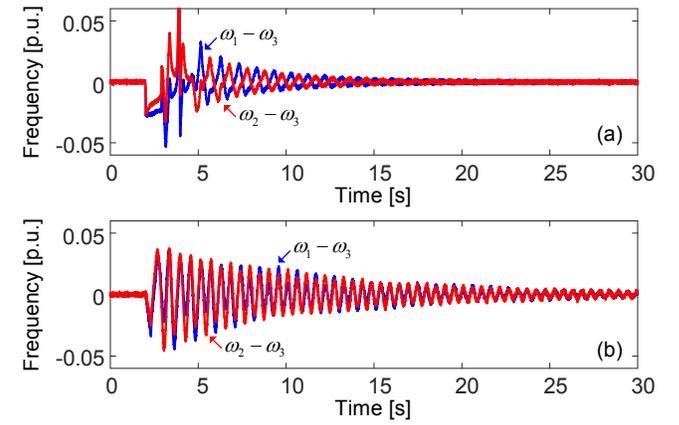

Fig. 14. Experimental result of the 9-bus system. (a) Hybrid system and (b) original system, where the fault lasts 0.90 s and 0.31 s, respectively.

## VI. CONCLUSION AND OUTLOOK

This paper addresses the modeling and analysis of transient stability for a hybrid power system with dynamics jointly dominated by two types of GFM devices. The transient synchronization behavior is equivalently described by a second-order equivalent motion equation. Based on this, two im-portant characteristics in the hybrid system synchronization behavior are discovered. Firstly, the damping characteristics are much enhanced (while the inertia remains unchanged), leading to a significant impact on the transient stability boundary. Secondly, the post-fault initial state is impacted by the droop frequency jump, bringing about a new factor that should be addressed in transient stability assessment. These two characteristics are identified as significant differences from the conventional power system. Regarding the droop-controlled system, it is explicitly indicated that it has an entirely different stability mechanism and stability boundary due to the overdamped and inertialess property.

In light of the findings in this study, a few enlightenments to future research can be provided. Specifically, attention should be paid to the transient stability of multi-machine multi-inverter hybrid power systems. In this regard, the damping effect should no longer be neglected when performing TSA, in which a less conservative TSA approach is required. The frequency jump should also be considered in the determination of CCA and CCT. This brings a new reminder to the design of relay protection. Moreover, making coordinated use of inertia emulation and droop control mixed resources is a new direction to address various types of stability in future power systems.

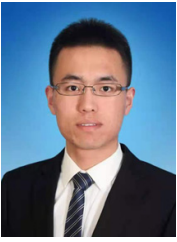

**Xiuqiang He** (S'17–M'21) received the B.S. degree in automation, in 2016, and the Ph.D. degree in control science and engineering, in 2021, from Tsinghua University, Beijing, China. He will join as a postdoctoral researcher the Automatic Control Laboratory, ETH Zürich, Switzerland.

His current research interests include stability issues of future power systems and generic modeling of renewable energy resources for power system dynamic studies.

Dr. He was the recipient of the Beijing Outstanding Graduates Award, the Outstanding Doctoral Dissertation Award of Tsinghua University, and the IEEE Transactions on Sustainable Energy Outstanding Reviewer Award in 2019. He is a member of IEC SC 8A WG 8 (Modeling of renewable energy generation for power system dynamic analysis).

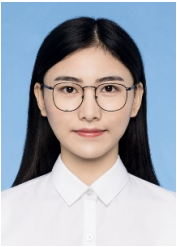

**Sisi Pan** received the B.E. degree, in 2018, from the Department of Electrical Engineering, Yangzhou University, Yangzhou, China, where she is currently working toward the MA.Eng. degree.

Her current research interests include hybrid real-time simulation for dynamic studies of modern power systems.

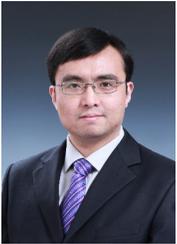

**Hua Geng** (S'07–M'10–SM'14–F'20) received the B.S. degree in electrical engineering from Huazhong University of Science and Technology, Wuhan, China, in 2003 and the Ph.D. degree in control theory and application from Tsinghua University, Beijing, China, in 2008. From 2008 to 2010, he was a Postdoctoral Research Fellow with the Department of Electrical and Computer Engineering, Ryerson University, Toronto, ON, Canada. He joined the Department of Automation Tsinghua University in June 2010 and is currently a full professor.

He has authored more than 170 technical publications and holds more than 20 issued Chinese or US patents. His current research interests include advanced control on power electronics and renewable energy conversion systems.

Dr. Geng was granted the second prize of the National Science and Technology Progress Award. He is the Editor of the IEEE TRANSACTIONS ON ENERGY CONVERSION and IEEE TRANSACTIONS ON SUSTAINABLE ENERGY, an Associate Editor for the IEEE TRANSACTIONS ON INDUSTRY APPLICATIONS, *IET Renewable Power Generation*, and *Control Engineering Practice*. He was the General Co-Chair, Track Chair, and Session Chair of various IEEE conferences. He is an IET Fellow, a Convener of IEC SC 8A WG 8, the Standing Director of China Power Supply Society, and the Vice Chair of IEEE IAS Beijing Chapter.